# Mechanical modeling of superelastic tensegrity braces for earthquake-proof structures

Fernando Fraternali

Department of Civil Engineering, University of Salerno

84084 Fisciano(SA), Italy

*f.fraternali@unisa.it*

Filipe Santos

Department of Civil Engineering, Universidade NOVA de Lisboa

2829-516 Lisboa, Caparica Portugal *fpas@fct.unl.pt*

October 2, 2019

## 1 Introduction

There is a growing scientific interest toward the design and engineering of nonlinear acoustic and mechanical metamaterials with tunable properties and advanced functionalities that are not found in conventional materials [1, 2, 3, 4, 5, 6, 7]. Seismic metamaterials are peculiar; they protect buildings and infrastructures from seismic waves, either by creating a shadow zone around the structure to be protected [8, 9, 10, 11] or by applying seismic isolation devices based on lattice metamaterials at the foundation level [12, 13]. Nonlinear metamaterials are progressively emerging as structured materials that can tune their responses to the level of the applied stress/strain or the amplitude of traveling waves [3, 14, 15, 16, 17]. Particularly interesting is the class of nonlinear metamaterials with tensegrity architecture: their mechanical behavior can be effectively adjusted by acting on internal and external prestress [18, 19, 20, 21, 22, 23, 24, 25, 26, 27, 28, 29], as well as the usual controls of geometry, topology, and mechanical properties of the members [30].

Dissipative bracing systems provide effective tools to passively dissipate energy in structures that are subject to severe earthquakes [31, 32, 33]. Such systems behave as nonlinear springs, interacting with the structure, and are able to dissipate energy in the form of inelastic deformation, typically under alternate tension/compression loading. The current technology is mainly based on the use of steel elements that yield when subjected to significant relative displacements or friction mechanism. Two major limitations of these devices are the need for replacement after yielding occurs as well as the bulky dimensions needed to prevent buckling. It should also be considered that the dissipation offered by structural braces is proportional to the interstory drifts of the braced buildings [34], which can be relatively low (of the order of magnitude of 1-2 % of the interstory height, or lower) [35, 36]. To mitigate against these drawbacks, the use of displacement amplification dampers and superelastic Shape Memory Alloy (SMA) elements within bracing systems have been intensively investigated in recent years [37, 32, 33, 38, 39, 40, 41, 42, 43, 44, 45].

The present paper deals with an unprecedented use of a a tensegrity structure equipped with SMA cables as an anti-seismic bracing system (hereafter referred to as, 'SMAD brace'). It consists of a D-bar system [18], i.e., a planar tensegrity structure composed of four bars forming a rhombus, which is internally stabilized through the insertion of two perpendicular SMA cables (or strings). Recently, D-bar systems have been recently actively investigated as effective absorption devices [46], due to their unique ability to combine a lightweight design with marked energy storage within the strings. The SMAD brace studied here is able to markedly amplify the applied longitudinal displacement in the transverse direction due to its geometric nonlinearities, allowing the transverse SMA cable to experience marked axial strain. The structure is thus able to dissipate a large amount of energy through its superelastic response, without suffering appreciable permanent deformations [47]. The remarkable displacement amplification property of the SMAD brace is magnified when the system assumes a compact and tapered shape, which is particularly convenient for the design of noninvasive bracing systems. Its geometry almost coincides with that of the scissor-jack damper studied in [37], the difference being that the SMAD brace includes a longitudinal SMA string (not present in the scissor-jack damper) and replaces the viscous damper with an SMA cable. Given the proper design, the displacement amplification property of the SMAD device can easily surpass that exhibited by the more conventional diagonal-, chrevron-, and toggle-brace dampers used in the seismic design of buildings. It can be profitably employed to form novel seismic metamaterials that exhibit an unconventional mechanical response that is mainly derived from the geometry of their internal structure [2, 3]. The illustration of the mechanics of the SMAD brace is given in Section 2, while its application for the reinforcement of anti-seismic structures is presented in Section 3. This section also includes a comparative study between the superelastic energy dissipation capacities of different SMAD bracing systems under seismic loading and the viscous energy dissipation capacity of a scissor-jack damper [37]. The use of SMAD braces as building blocks for novel seismic metamaterials with tensegrity architecture is outlined in Section 4. The main conclusions of the present study and suggestions for future work are presented in Section 5.

## 2 Mechanical modeling of SMAD braces

Let us consider the SMAD brace illustrated in Fig. 1, which depicts the reference configuration of such a structure under zero external force (solid lines), and its deformed shape under the action of a vertical force *P* applied to the top vertex (dashed lines). We assume that the top vertex is constrained by a vertical roller, while the bottom vertex is constrained by a fixed hinge. The following sections illustrate the adopted mechanical model of the SMAD brace (Sect. 2.1), and the associated constitutive equations of the SMA wires (Sect. 2.2).

## 2.1 Members forces, mass reduction and displacement amplification properties

Throughout the paper, we admit that the SMAD brace is under a zero state of self-stress in the reference configuration, which we identify with the placement of the structure under the action of a zero external force $P$. Such state of prestress, whose study is beyond the scope of the present work, would follow, e.g., from a balanced pretension of the SMA cables [18]. Due to the symmetry of the SMAD brace and the examined loading condition, it is easy to recognize that all the bars of such a structure carry an equal axial force $N$, which we assume positive in compression. The longitudinal string instead carries a vertical tensile force $Y$, while the transverse string carries a horizontal tensile force $X$. By solving the equilibrium equations of the top and the side nodes in the deformed configuration, we obtain the following expressions of the members forces in equilibrium conditions (Fig. 1)

$$N = X/\cos\hat{\theta}, \qquad P = 2N\sin\hat{\theta} - Y = X\tan\hat{\theta} - Y \qquad (1)$$

where $\hat{\theta}$ denotes the angle formed by the bars and the horizontal string in the deformed configuration (we let $\theta$ denote the same angle in the reference configuration, cf. Fig. 1).

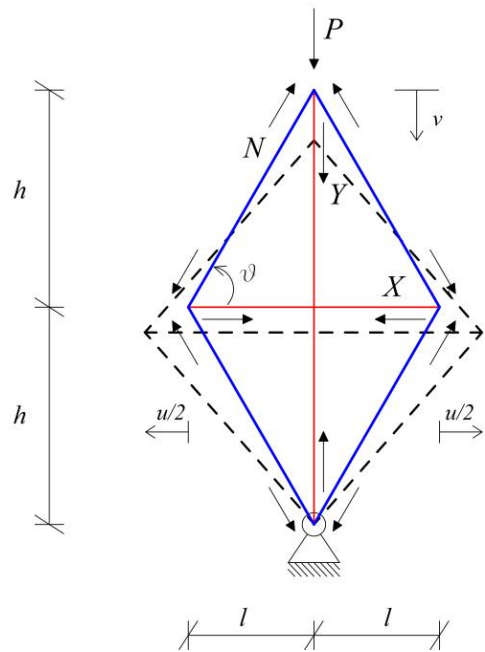

Figure 1. Illustration of the reference (solid lines) and deformed (dashed lines) configurations of the SMAD brace.

It is instructive to compare the buckling response of the SMAD brace with that of a straight beam with length $2h$, which is made out of the same material. Upon designing such structures so that they exhibit the same buckling load, and assuming that the strings go slack in compression, it is not difficult to show that the mass $m_1$ of the SMAD brace relates to the mass $m_0$ of the straight beam through the following relation (cf. [18], Par. 3.3)

$$m_1/m_0 = (2\sin^5\theta)^{-\frac{1}{2}} \tag{2}$$

This formula returns a $m_1/m_0$ ratio smaller than one when it results $\theta$ > 60.5 degrees, and one notices that it results $m_1/m_0 \approx 0.73$ when $\theta$ = 80 degrees (Fig. 2)). Such a remarkable result proves that a SMAD brace with tapered profile ($\theta \gg 60.5$ degrees) exhibits a significantly greater buckling load, as compared to a straight beam of equal mass.

Lets now assume that, during any arbitrary transformation of the structure, the bars behave as rigid bodies, i.e., it results

$$\left(h - \frac{v}{2}\right)^2 + \left(l + \frac{u}{2}\right)^2 = b^2 = \text{const} \tag{3}$$

where $b$ denotes the length of the bars, and $u$ and $v$ denote the longitudinal and the transverse displacements of the structure (cf. Fig. 1). Such an assumption is commonly accepted when dealing with tensegrity structures such that the axial stiffness of the bars is markedly greater than that of the strings (see, e.g., [18]). Eqn. (3) can be used to relate the transverse displacement $u$ to the longitudinal displacement $v$. Upon parameterizing the displacement $v$ by the bar length $b$, so that it results $v = \beta\, b$, it is not difficult to show that Eqn. (3) admits the following (feasible) solution

$$u/v = \left(\sqrt{4\cos^2(\theta) - \beta(\beta - 4\sin(\theta))} - 2\cos(\theta)\right)/\beta \tag{4}$$

In the limiting case with $\beta \to 0$, which corresponds to assuming infinitesimal displacements from the reference configuration, it is easily shown that Eqn. (4) predicts $u/v = \tan\theta$. Fig. 2 illustrates the variation of $u/v$ with $\theta$ for different values of $\beta$.

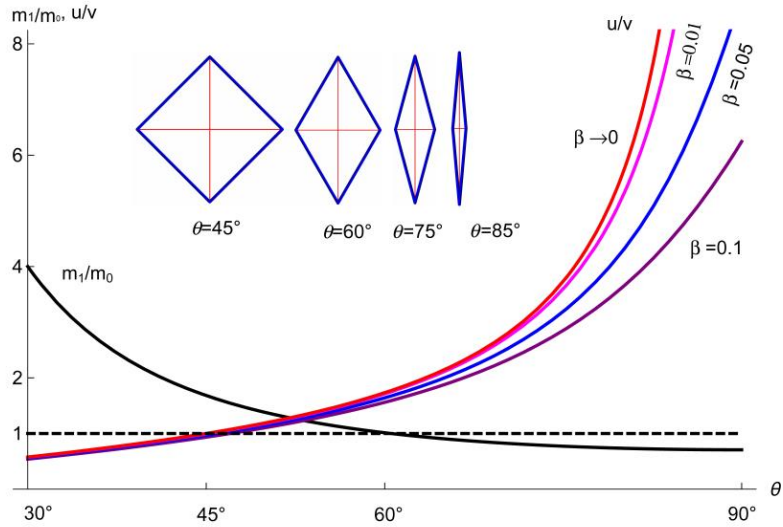

Figure 2. Mass reduction and displacement amplification factors of the SMAD brace.

Referring to the small displacement regime, the results in Fig. 2 highlight that the *u/v* ratio tends to infinity as *θ* approaches 90 degrees, and one observes that it results *u/v* = 5.67 already for *θ* = 80 degrees. Upon setting *β* = 0.01 and *θ* = 80 degrees, one instead gets a slightly lower value of the *u/v* ratio, which is nevertheless again significantly large (*u/v* = 5.26).

## 2.2   Constitutive response of the SMA cables

The constitutive response of SMAs shows a high level of complexity, being closely connected with the austenitic (A) - martensitic (M) phase transformations occurring within the material, and the thermodynamics underlying such processes (refer, e.g., to [47, 48, 49, 50] and references therein). The present work describes the superelastic response of the Nickel-Titanium (NiTi) cables forming the SMAD brace through an internal variable model with explicit phase transformation kinetics. Use is made of the Tanaka model [47], which proved to lead to accurate results when modeling one-dimensional elements in structural engineering applications [51, 52]. The adopted constitutive law for the stress *σ* acting in the generic cable is the following

$$\sigma = [\xi E_M + (1 - \xi)E_A](\varepsilon - \varepsilon_L \xi) + \Theta(T - T_0) \tag{5}$$

where *ε* is the axial strain; $\varepsilon_L$ is the maximum transformation strain; *ξ* is the internal variable describing the martensite fraction of the material; $E_A$ and $E_M$ are the austenitic and martensitic elastic moduli, respectively; *T* is the cable temperature; Θ is the thermal coefficient of expansion; and $T_0$ is the temperature at which the thermal strain is zero (see, e.g., [51] for a detailed illustration of the present model). The evolution laws for the direct (A)→(M) and the inverse (M)→(A) phase transformations are written as follows

$$\xi^{AM} = 1 - \exp[a_M(M_s - T) + b_M \sigma], \qquad \xi^{MA} = \exp[a_A(A_s - T) + b_A \sigma] \tag{6}$$

Said $A_s$, $A_f$, $M_s$ and $M_f$ the austenite start temperature, austenite finish temperature, martensite start temperature, and martensite finish temperature, respectively, the constants $a_A$, $b_A$, $a_M$, and $b_M$ appearing in Eqn. (6) are computed through

$$a_M = -\frac{2\ln 10}{M_s - M_f}, \quad b_M = \frac{a_M}{C_M} \quad (A \to M), \quad a_A = \frac{2\ln 10}{A_f - A_s}, \quad b_A = \frac{a_A}{C_A} \quad (M \to A) \tag{7}$$

$C_A$ and $C_M$ denoting the Clausius-Clapeyron coefficients measuring the slope of the stress-temperature plot at characteristic temperatures [51]. While the quasi-static loading of a SMA cable can be approximated as a nearly isothermal process, the fast dynamic cycling that occurs, e.g., during seismic excitations needs to be modeled by adding a heat transfer equation to the mechanical and kinetic transformation laws given above. By making use of the model of convection heat transfer from surfaces presented in [53], we write

$$-\varrho c V \frac{dT}{dt} = \overline{h} A [T - T_f] - q_{gen} V \qquad \text{with } T(0) = T_f \tag{8}$$

where *V* is the volume of the cable: *A* is the area of the lateral surface; ϱ is the material density; *h* is the heat transfer coefficient, and $q_{gen}$ is the power generated per unit volume, which is computed as follows

$$q_{gen} = c_L \varrho \frac{d\xi}{dt} + \frac{dW}{dt} \tag{9}$$

The first term on the rhs of Eqn. (9) is related to the time derivative of the martensite fraction $\xi$, assuming constant latent heat $c_L$. The second term is the time derivative of the energy $W$ dissipated within the material due to the hysteretic stress-strain response of the cable (internal friction).

### 2.2.1 Experimental validation

We conducted an experimental validation of the constitutive model illustrated in the previous section by performing tensile tests on a NiTi wire with 0.406 mm diameter through a Zwick/Roell Z050 electro-mechanical testing machine. Such tests were performed in quasi-static loading conditions at the ambient temperature of 20° C, using a strain rate of $2.5 \times 10^{-5}$ s$^{-1}$. After applying 20 loading-unloading preconditioning cycles, the tested wire was subject to tensile tests up to different maximum axial strains $\varepsilon_{max}$, ranging from 2% to 6%. These tests were numerically simulated using the constitutive parameters illustrated in Tab. 1. Figure 3 shows a comparison between the experimental results and the numerical simulations of the tensile tests, which highlights a good theory vs. experiment matching. It is worth noting that very small residual strains are experimentally observed at unloading (not greater than 0.03 $\varepsilon_{max}$). The latter are neglected by the constitutive model presented in the previous section, due to their minor relevance.

Table 1. Constitutive parameters of the adopted SMA model.

| Parameter | Value | Parameter | Value |
|---|---|---|---|
| $E_A$ | 35 GPa | $M_s$ | -45 °C |
| $E_M$ | 20 GPa | $M_f$ | -65 °C |
| $\Theta$ | 0.38 MPa °C$^{-1}$ | $A_s$ | 5 °C |
| $\varepsilon_L$ | 5.0 % | $A_f$ | 10 °C |
| $C_M$ | 6.5 MPa °C$^{-1}$ | $c$ | 550 J kg$^{-1}$ °C$^{-1}$ |
| $C_A$ | 6.5 MPa °C$^{-1}$ | $c_L$ | 12914 J kg$^{-1}$ |
| $\%$ | 6.5 g/cm$^3$ | | 35 W m$^{-2}$ °C$^{-1}$ |
| | | $\bar{h}$ | |

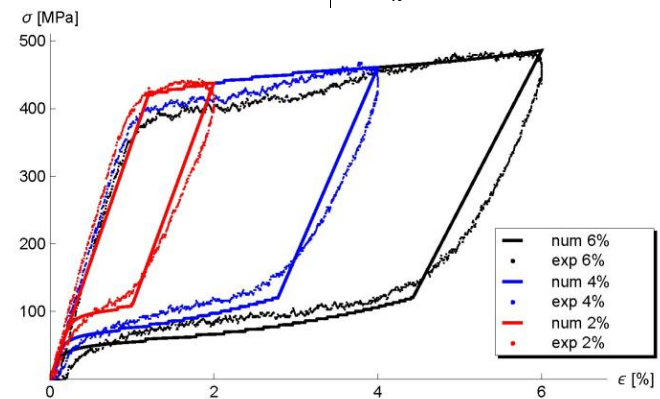

Figure 3. Comparison between the results of tensile tests on a NiTi wire and numerical simulations using the material constants in Tab. 1.

## 3 Use of SMAD braces in anti-seismic structures

Let us refer to the frame structure illustrated in Fig. 4, which is equipped with hinged beamto-column and column-to-foundation connections. The frame is seismically reinforced through the insertion of two symmetric SMAD braces, which allow the structure to resist horizontal loads $F$ produced by seismic movements of the foundation. When the force $F$ is directed as in Fig. 4, the right SMAD brace is loaded by a compressive force $P_1$, while the left brace is loaded by a tensile force $P_1$. We denote the horizontal displacement of the frame by $u_f$, and the insertion angles of the SMAD braces by $\psi$ (Fig. 4). The geometric compatibility between the frame and the brace requires

$$v = u_f \cos\psi, \quad h = \frac{H}{2\sin\psi}, \quad b = \frac{H}{2\sin\psi\sin\theta}, \quad \beta = \frac{v}{b} = 2\,\delta\,\sin\psi\cos\psi\sin\theta \quad (10)$$

where $\delta = u_f/H$ denotes the drift ratio of the frame. Making use of Eqns. (10) into Eq. (3), we finally get the following expression of the lateral amplification factor $u/u_F$ of the SMAD brace

$$f = u/u_F = \left[\csc(\psi)\csc(\theta)\left(\sqrt{\delta\sin(\psi)\cos(\psi)\sin^2(\theta)(2-\delta\sin(\psi)\cos(\psi))+\cos^2(\theta)} - \cos(\theta)\right)\right] \quad (11)$$

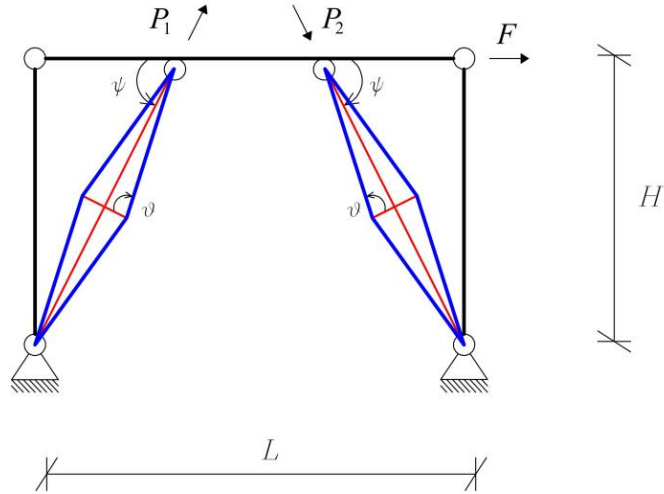

Figure 4. Frame structure strengthened with symmetric SMAD braces.

It is easy to verify that Eq. (11) reduces to $f = \cos\psi \tan\theta$ in the small displacement regime ($\delta \to 0$). One also notes that large values of $\psi$ (i.e., $\psi \to \pi/2$) reduce the displacement amplification factor given by Eq. (3). Fig. 5 graphically illustrates the predictions of Eqn. (11) for different values of $\psi$ and $\theta$, assuming $\delta = 1\%$. Such value of the drift ratio corresponds to an average value of the thresholds dictated by the international standards for the seismic design of buildings, in order to avoid structural collapse (refer, e.g., to Eurocode 8 [54]).

Let us now compare the values of $f$ predicted by Eq. (11) with the displacement amplification factors offered by devices that make use of commercial viscous dampers. Referring to the benchmark examples presented in [37], we note that $f$ assumes values equal to 0.80, 1.00, 2.16, 3.19 and 2.51 for diagonal, chevron, scissor-jack, upper toggle and reverse toggles dampers, respectively (see Fig. 1 of Ref. [37] for a graphical illustration of such devices). We already remarked that the SMAD brace has the same geometry of the scissor-jack damper, even if it makes use of the superasticty of SMA cables instead of viscous effects to dissipate energy. The results shown in Fig. 5 highlight that it possible to achieve values of *f* equal to 3 or larger, through an accurate design of the aspect angles $\psi$ and $\theta$ of the SMAD brace. For a given value of $\psi$, which may be dictated by architectural or technical needs, one can suitably taper the SMAD brace by increasing $\theta$ up to values close to 90 degrees. This is possible thanks to the use of SMA cables in the SMAD brace, as opposed to the case of the scissor-jack device [37], where the transverse dimension *l* needs to be large enough in order to accommodate a viscous damper within the brace. The insets plots in Fig. 5 provide examples of geometric designs of the SMAD brace that lead to achieve *f* = 3, for $\delta$ = 1%. The achievement of larger values of *f* is also possible (cf. Fig. 5). Using Eqn. (11) and again assuming $\delta$ = 1%, we observe, e.g., that a displacement amplification factor *f* = 5 is obtained when using the following couples of values of the SMAD aspect angles: ($\psi$ = 45 deg, $\theta$ = 83 deg); ($\psi$ = 60 deg, $\theta$ = 85 deg); ($\psi$ = 70 deg, $\theta$ = 87 deg).

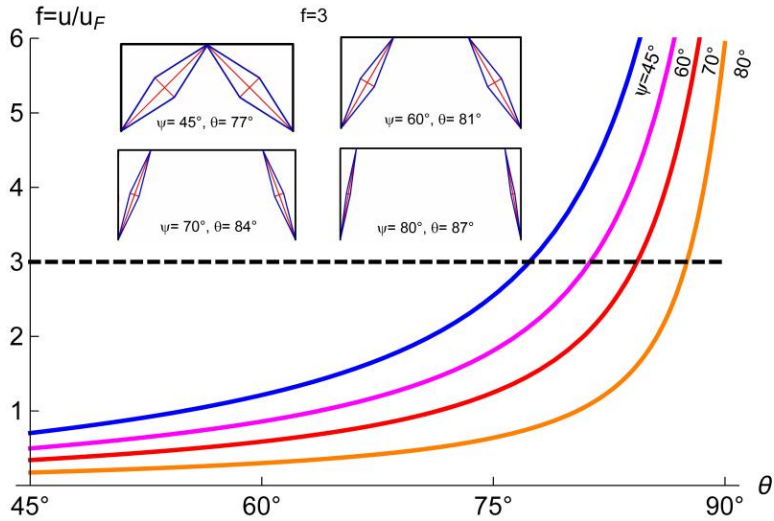

Figure 5. Lateral displacement amplification ratio of the SMAD brace for $\delta$ =1% and different values of the aspect angles $\psi$ and $\theta$.

In order to evaluate the energy dissipation capacity of real-life implementations of SMAD braces, we hereafter examine the reinforcement of the model frame structure analyzed in [37], which was tested on an earthquake simulator at the University of Buffalo. The analyzed frame exhibits bay length $L$ = 2540 mm, height $H$ = 1927 mm, a steel wide flange beam W8×21, and wide flange columns W8×24. Such a structure was strengthened with a scissor-jack damper (SJD) in [37], which was preferred to diagonal, chevron, upper toggle and reverse toggles dampers due to its enhanced displacement magnification property and easy of implementation. The SJD device analyzed in [37] features aspect angles $\psi$ = 70 deg and $\theta$ = 81 deg (notice that Ref. [37] labels $\theta$ the complementary of the aspect angle $\theta$ introduced in the present work). It is framed by square hollow section profiles TS 2×2×1/4". The fluid viscous damper has 273 mm length, diameter of 44.5 mm, stroke of ±28.6 mm, and is described by the response law $X = Cu^{\cdot \alpha}$, with $C$ = 137.3 N (s/mm)$^\alpha$, and $\alpha$ = 0.76 (we refer the reader to Ref. [37] for further technical details about the structure under examination). While the frame experimentally studied in [37] features coupled simply-rigid beam-to-column connections, our SMAD design assumes all simply connected connections, in agreement with the schematic shown in Fig. 4. We account for several geometric designs of the two SMAD braces, making use of the TS 2×2×1/4' profiles of the scissor-jack design for the bars, and a transverse cable consisting of a SMA strand formed by 12+6+1 wires [55], each of which has 2 mm diameter. The use of two symmetric SMAD braces without longitudinal cables allows us to obtain a symmetric lateral force $F$ vs. lateral displacement $u_f$ response of the frame, which engages only the right brace for rightward directed forces $F$, and the only the left brace for leftward directed forces $F$. Such a design leads us to easily predict the response of the frame to lateral forces, which is ruled by the transverse SMA cables. It can be easily generalized to include longitudinal SMA cables within the D-bar units, whose presence will cause both the braces to be engaged, under arbitrarily directed forces $F$.

Fig. 6 displays the $X$ vs. $u$ responses of different SMAD bracing reinforcements of the frame model under consideration, which were obtained by making use of the constitutive parameters described in Tab. 1. The SMAD responses are compared with the experimental response reported in [37] for the SJD. The latter refers to the third cycle of a displacement loading at the frequency of 4 Hz. Our mechanical modeling of the SMAD response assumes cycling loading with the achievement of a maximum lateral displacement $u_F$ = 6 mm (as in Fig.8 of [37]). The transverse force-displacement laws of the SMAD systems reported in Fig. 6 exhibit the characteristic flagshaped hysteretic response that is reported in the literature for SMA cables [47, 48, 49, 50, 51]. One observes that the examined SMAD braces develop maximum forces more than three times larger than the maximum force exhibited by the SJD ($\approx$ 7.7 kN). The highest cable force $X$ is observed in the SMAD brace featuring ($\psi$ = 80 deg. $\theta$ = 85 deg), which is equal to 33 kN. On considering that it results $N = X/\cos\hat{\theta}$ (cf. Eqn. (1)), we predict a peak force in the struts equal to 378 kN. Since the corresponding buckling load $N_{cr}$ amounts to 666 kN, we are led conclude that such a structure structure exhibit a rather high safety factor against buckling of the struts ($N_{max}/N_{cr}$ = 0.57), as well as all the other SMAD braces examined in Fig. 6. We wish to remark that the length $l$ of the SMA strand in the ($\psi$ = 80 deg. $\theta$ = 85 deg) design is equal to 171 mm, i.e. 0.63 times the length of the damper employed in the scissor-jack brace (273 mm).

The energies dissipated by the ($\psi$ = 80 deg. $\theta$ = 85 deg), ($\psi$ = 70 deg. $\theta$ = 82.5 deg), and ($\psi$ = 70 deg. $\theta$ = 80 deg) SMAD braces, through superelastic response, are 33 %, 68%, and 2% greater than the energy dissipated by the SJD (Fig. 6). We remark the dramatical increase of the dissipated energy offered by the ($\psi$ = 70 deg. $\theta$ = 82.5 deg) SMAD braces, over the SJD, by noting that such a bracing system exhibits geometry very similar to the examined scissor-jack device ($\psi$ = 70 deg. $\theta$ = 81 deg) [37]. It is also worth to mention the remarkable energy dissipation gain that is obtained through the extremely compact SMAD braces with ($\psi$ = 80 deg. $\theta$ = 85 deg), always comparing the energy performance of the SMAD braces with that of the SJD.

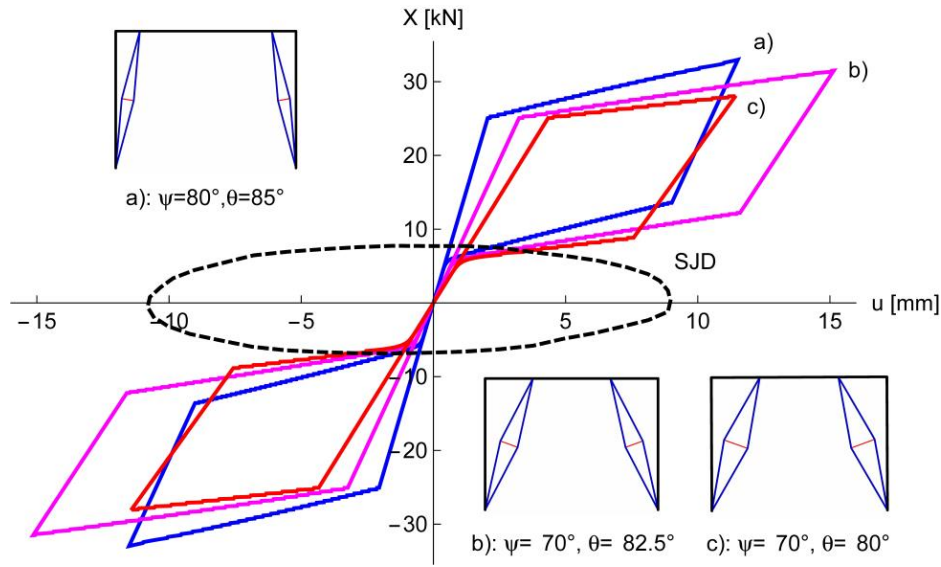

Figure 6. Force-displacement responses of several bracing systems reinforcing a model frame [37].

## 4 Seismic metamaterials using SMAD building blocks

The currently accepted definition of seismic metamaterials refers to discrete structures forming shields that protect buildings from seismic waves, or lattice structures serving as unconventional seismic isolation devices [12, 13]. We here propose structural uses of SMAD braces that allows us to stretch such a definition, with the aim of including bracing systems which make use of SMAD building blocks, and exhibit displacement amplification capacity and energy dissipation properties surpassing those of conventional anti-seismic braces. Fig. 7 illustrates a parade of tensegrity braces which tessellate SMAD units, and are inserted within the bays of seismically resistant frame structures. Making use of fractal geometry concepts, such systems can be designed through selfsimilar divisions of the basic module, in order to enhance the energy dissipation capacity, as well as the buckling resistance of the bracing system [18]. The buckling of the units can be also prevented through internal pre-tension of the braces via internal self-stress, and contrasting them against the elements of the served structure (external pre-tension). Such extremely lightweight and noninvasive bracing systems will be easily inserted around window frames or holes, or assembled as self-contained architectural components (e.g., window frames) to be used in retrofit and upgrade interventions. Since SMA cables are not affected by relevant permanent deformations under cyclic loading, the SMAD metamaterials will also exhibit a convenient re-centering ability after earthquakes [32, 33].

## 5 Concluding remarks

We have presented a novel use of tensegrity systems to form superelastic braces for seismically resistant structures, which are aimed at preventing or minimizing structural damage during seismic excitations. The analyzed bracing devices consist of D-bar structures equipped with SMA cables, which exhibit extreme displacement amplification properties. Likewise, they have a lightweight design to mitigate against buckling, high energy dissipation capacity, and strong re-centering ability. The maximum performances of the SMAD brace in terms of displacement magnification are achieved when using a tapered profile of the structure, which is convenient for concealing the brace for architectural purposes (Sections 2-3). We have also shown that a suitable design of SMAD braces gives rise to an energy dissipation capacity that markedly surpasses that offered by more conventional bracing systems based on fluid viscous dampers (Section 3). The main advantages of SMAD braces over viscous dampers derive from the use of SMA cables in place of more widespread dissipation devices, the increase of the buckling resistance with the tapering of the structure, their re-centering capacity, and their displacement amplification properties being based on the geometry of the system rather than the chemical nature of the material. The latter feature suggests that SMAD braces can be used as building blocks for novel seismic metamaterials (see Section 4), and as part of additive manufacturing techniques used to construct the physical models of these systems [56, 57, 58, 59]. It is also worth observing that SMAD braces are less exposed to reductions in displacement magnification due to dynamic loading [37]. This is due to the non-viscous nature of the energy dissipated by the SMA cables. Similarly, it should be mentioned that fluid viscous dampers can be subject to fluid leakage [60], which does not affect SMAD braces.

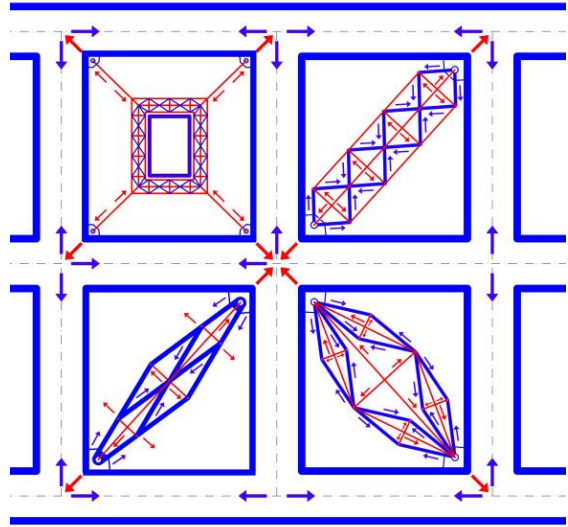

Figure 7. Tensegrity braces built out of SMAD units.

Future directions for the present research are manifold. An initial extension is to investigate the effects produced by internal and external states of prestress on the mechanics of SMAD braces. These effects are expected to symmetrize the structure's response under reverse loading and to engage both the longitudinal and the transverse cables under arbitrarily directed lateral forces. A second line of future research regards the development of seismic metamaterials tessellating SMAD units over one or multiple directions in space, with recourse to self-similar design approaches (Section 4). The experimental verification of the response of SMAD braces under seismic loading also requires attention. We plan to carry out this work using a comparative analysis that matches the response of SMAD braces with that of more conventional bracing systems, using fast dynamic loading and full-scale or reduced-scale physical models.

## Acknowledgements

This is a pre-print of the article published into Extreme Mechanics Letters with the same title, ISSN:2352-4316, doi: https://doi.org/10.1016/j.eml.2019.100578

FF gratefully acknowledges financial support from the Italian Ministry of Education, University and Research (MIUR) under the PRIN 2017 Grant 'Multiscale Innovative Materials and Structures'.

## References

[1] J. Christensen, M. Kadic, O. Kraft, M. Wegener, 2015, Vibrant times for mechanical metamaterials, MRS Commun, 5(3) (2015) 453-462.

[2] S.A. Cummer, J. Christensen, A. Alu, Controlling sound with acoustic metamaterials, Nat Rev Materials, 1, 16001 (2016).

[3] K. Bertoldi, V. Vitelli, J. Christensen, M. Van Hecke, Flexible mechanical metamaterials, Nat Rev Materials, 2, 17066 (2017).

[4] J. Rys, S. Steenhusen, C. Schumacher, C. Cronauer, C. Daraio, Locally addressable material properties in 3D micro-architectures, Extreme Mechanics Letters (2019).

[5] O. Bilal, A. Foehr, C. Daraio, Observation of trampoline phenomena in 3D-printed metamaterial plates, Extreme Mechanics Letters, 15 (2017).

[6] J.J. do Rosario, J.B. Berger, E.T. Lilleodden, R.M. McMeeking, G.A. Schneider, The stiffness and strength of metamaterials based on the inverse opal architecture, Extreme Mech. Lett., 12 (2017) 86-96

[7] Y. Tang, J. Yin, Design of cut unit geometry in hierarchical kirigami-based auxetic metamaterials for high stretchability and compressibility, Extreme Mech. Lett., 12 (2017) 77-85

[8] S. Kroedel, N. Thome, C. Daraio, Wide band-gap seismic metastructures, Extreme Mech. Lett. 4 (2015) 111-117.

[9] A. Colombi, D. Colquitt, P. Roux, S. Guenneau, R.V. Craster, A seismic metamaterial: The resonant metawedge, Sci Rep, 6:27717 (2016).

[10] Q. Du, Y. Zeng, G. Huang, H. Yang, Elastic metamaterial-based seismic shield for both lamb and surface waves, AIP Adv, 7, 075015 (2017).

[11] P. Roux, D. Bindi, T. Boxberger, A. Colombi, F. Cotton, I. Douste-Bacque, I. Pondaven, Toward seismic metamaterials: The METAFORET project, SeismoL. Res Lett, 89(2A) (2018) 582-593.

[12] F. Fraternali, A. Amendola, Mechanical modeling of innovative metamaterials alternating pentamode lattices and confinement plates, J Mech Phys Solids 99, (2017) 259-271.

[13] F. Fraternali, A. Amendola, G. Benzoni, Innovative seismic isolation devices based on lattice materials: A review. Ingegneria Sismica: International Journal of Earthquake Engineering, 4 (2018) 93-113.

[14] M.I. Hussein, M.J. Leamy, M. Ruzzene, Dynamics of phononic materials and structures: Historical origins, recent progress, and future outlook, Appl Mech Rev, 66(4) (2014).

[15] R.K. Narisetti, M.J. Leamy, M. Ruzzene, A perturbation approach for predicting wave propagation in one-dimensional nonlinear periodic structures, J Vib Acoust, Transactions of the ASME, 132(3), 0310011-03100111 (2010).

[16] M.P. O'Donnell, P.M. Weaver, A. Pirrera, Can tailored non-linearity of hierarchical structures inform future material development? Extreme Mech. Lett., 7 (2016) 1–9

[17] J.T. Klein, E.G. Karpov, Bistability in thermomechanical metamaterials structured as threedimensional composite tetrahedra, Extreme Mech. Lett., 29 (2019) 100459

[18] R.E. Skelton, M.C. de Oliveira, Tensegrity Systems, Springer (2010).

[19] F. Fraternali, L. Senatore, C. Daraio, Solitary waves on tensegrity lattices, J Mech Phys Solids 60 (2012) 1137-1144.

[20] F. Fraternali, G. Carpentieri, A. Amendola, On the mechanical modeling of the extreme softening/stiffening response of axially loaded tensegrity prisms, J Mech Phys Solids, 74 (2014) 136-157.

[21] A. Amendola, G. Carpentieri, M. de Oliveira, R.E. Skelton, F. Fraternali, Experimental investigation of the softening-stiffening response of tensegrity prisms under compressive loading, Compos Struct 117 (2014) 234-243.

[22] F. Fraternali, G. Carpentieri, A. Amendola, R.E. Skelton, V.F. Nesterenko, Multiscale tunability of solitary wave dynamics in tensegrity metamaterials, Appl Phys Lett, 105, 201903 (2014).

[23] K. Pajunen, P. Johanns, P. Raj Kumar, J. Rimol, C. Daraio, Design and Impact Response of 3D-Printable Tensegrity-Inspired Structures, Materials & Design. 107966 (2019).

[24] A. Krushynska, A. Amendola, F. Bosia, C. Daraio, N. Pugno, F. Fraternali, Accordion-like metamaterials with tunable ultra-wide low-frequency band gaps, New Journal of Physics 20 (2018).

[25] J.J. Rimoli, R.K. Pal, Mechanical response of 3-dimensional tensegrity lattices, Compos. Part B-Eng. 115 (2017) 30-42.

[26] J.J. Rimoli, A reduced-order model for the dynamic and post-buckling behavior of tensegrity structures, Mech. Mater. 116 (2018) 146-157

[27] R.K. Pal, M. Ruzzene, J.J. Rimoli, Tunable wave propagation by varying prestrain in tensegrity-based periodic media, Extreme Mech. Lett. 22 (2018) 149-156

[28] H. Salahshoor, R.K. Pal, J.J. Rimol, Material symmetry phase transitions in three-dimensional tensegrity metamaterials, J. Mech. Phys. Solids 119 (2018) 382-399

[29] A. Amendola, A. Krushynska, C. Daraio, N.M. Pugno, F. Fraternali, Tuning frequency band gaps of tensegrity metamaterials with local and global prestress, Int. J. Solids Struct., 155 (2018) 47–56.

[30] A.G. Tilbert, S. Pellegrino, Review of form-finding methods for tensegrity structures, Int J Space Struct, 18 (2011) 209-223.

[31] M. D. Symans, F.A. Charney, A.S. Whittaker, M.C. Constantinou, C.A. Kircher, M.W. Johnson, R. J. McNamara, Energy Dissipation Systems for Seismic Applications: Current Practice and Recent Developments, J Struct Eng-Asce, 134 (2008), 3-21.

[32] C. Menna, F. Auricchio, D. Asprone, 2014. Applications of shape memory alloys in structural engineering, Shape memory alloy engineering. Boston: Butterworth-Heinemann, (2015) 369-403.

[33] W. Chang, Y. Araki, Use of shape-memory alloys in construction: A critical review, P I Civil Eng: Civil Engineering, 169(2) (2016) 87-95.

[34] Mathias, N., F. Ranaudo, M. Sarkisian, Mechanical amplification of relative movements in damped outriggers for wind and seismic response mitigation, International Journal of High-Rise Buildings 5 (1) (2016) 51-62.

[35] F. Barbagallo, M. Bosco, E.M. Marino, P.P. Rossi, Seismic design and performance of dual structures with BRBs and semi-rigid connections, Journal of Constructional Steel Research, 158 (2019) 306-316.

[36] Chou, C., C. Hsiao, Z. Chen, P. Chung, D. Pham, D., Seismic loading tests of full-scale twostory steel building frames with self-centering braces and buckling-restrained braces. Thin-Walled Structures, 140 (2019) 168-181.

[37] Şigaher, A.N., Constantinouu, M.C., Scissor-jack-damper energy dissipation system, Earthquake Spectra, 19(1) (2003) 133-158.

[38] C.W. Yang, R. DesRoches, R.T. Leon, Design and analysis of braced frames with shape memory alloy and energy-absorbing hybrid devices, Eng Struct, 32(2) (2010) 498-507.

[39] D.J. Miller, L.A. Fahnestock, M.R Eatherton, 2012. Development and experimental validation of a nickel-titanium shape memory alloy self-centering buckling-restrained brace, Eng Struct, 40 (2012) 288-298.

[40] B. Asgarian, N. Salari, B. Saadati, Application of intelligent passive devices based on shape memory alloys in seismic control of structures, Structures, 5 (2016) 161-169.

[41] M. Bosco, E.M. Marino, P.P. Rossi, A design procedure for pin-supported rocking bucklingrestrained braced frames, Earthq Eng Struct D, 47(14) (2018) 2840-2863.

[42] M. Liu, P. Zhou, H. Li, Novel self-centering negative stiffness damper based on combination of shape memory alloy and prepressed springs, J Aerospace Eng, 31(6) (2018).

[43] W. Wang, C. Fang, J. Liu, Large size superelastic SMA bars: Heat treatment strategy, mechanical property and seismic application, Adv Mater Res-Switz, 25(7) (2016).

[44] M. Dolce, D. Cardone, R. Marnetto, Implementation and testing of passive control devices based on shape memory alloys, Earthq Eng Struct D, 29(7) (2000) 945-68.

[45] S. Das, S.K. Mishra, Optimal performance of buildings isolated by shape-memory-alloy-rubberbearing under random earthquakes, Int J Comput Meth Eng Sci Mech, 15(3) (2014) 265-276.

[46] R. Goyal, E.A. Perrera Hernandez, R.E. Skelton Analytical study of tensegrity lattices for mass-efficient mechanical energy absorption, International Journal of Space Structures 0956059919845330 (2019).

[47] K. Tanaka, S. Kobayashi, Y. Sato, Thermomechanics of transformation pseudoelasticity and shape memory effect in alloys, International Journal of Plasticity, 2 (1986) 59–72.

[48] K. Tanaka, F. Nishimura, T. Hayashi, H. Tobushi, and C. Lexcellent, Phenomenological analysis on subloops and cyclic behavior in shape memory alloys under mechanical and/or thermal loads, Mechanics of Materials, 19 (1995) 281–292.

[49] C. Liang, C.A. Rogers. One-Dimensional Thermomechanical Constitutive Relations for ShapeMemory Materials, Journal of Intelligent Systems and Structures, 1 (1990) 207–234.

[50] F. Auricchio and E. Sacco, A one-dimensional model for superelastic shape-memory alloys with different elastic properties between austenite and martensite, International Journal of Non-Linear Mechanics, 32(6) (1997) 1101–1114.

[51] F.P. Amarante dos Santos, C.Cismaşiu, Comparison Between Two SMA Constitutive Models for Seismic Applications. Journal of Vibration and Control, 16(6) (2010) 897–914.

[52] C. Cismasiu and F.P. Amarante dos Santos, Numerical simulation of superelastic shape memory alloys subjected to dynamic loads, Smart Materials and Structures, 17(2) (2008) 25–36.

[53] L.C. Thomas, Heat Transfer, Prentice-Hall, Inc. (1992).

[54] EN 1998, Eurocode 8: Design of structures for earthquake resistance, Brussels: Belgium (2004).

[55] V. Mercuri, Shape memory alloy strands: Conventional 3D FEM modeling and simplified models, MSc Thesis, University of Pavi, (2013/2014).

[56] A. Amendola, E.H. Nava, R. Goodall, I. Todd, R.E. Skelton, F. Fraternali, On the additive manufacturing, post-tensioning and testing of bi-material tensegrity structures, Compos. Struct. 131 (2015) 66-71.

[57] T. Tacone-Dejan, D. Mohr, Elastically-isotropic elementary cubic lattices composed of tailored hollow beams, Extreme Mech. Lett. 22 (2018) 13–18

[58] J. Mueller, K. Shera, Stepwise graded struts for maximizing energy absorption in lattices, Extreme Mech. Lett. 25 (2018) 7–15

[59] X. Tana, S. Chena, B. Wanga, S. Zhua, L. Wua,Y. Sun, Design, fabrication, and characterization of multistable mechanical metamaterials for trapping energy, Extreme Mech. Lett. 28 (2019) 8–21

[60] Z.D. Xu, Y.Q. Guo, J.T. Zhu, F.H. Xu, Intelligent Vibration Control in Civil Engineering Structures, Academic Press, New York (2016).